\documentclass[journal]{IEEEtran}

\usepackage{xcolor,soul,framed} 
\usepackage[justification=centering]{caption}

\colorlet{shadecolor}{yellow}
\usepackage[pdftex]{graphicx}
\graphicspath{{../pdf/}{../jpeg/}}
\DeclareGraphicsExtensions{.pdf,.jpeg,.png}

\usepackage{units}
\usepackage[cmex10]{amsmath}
\usepackage{array}
\usepackage{mdwmath}
\usepackage{mdwtab}
\usepackage{eqparbox}
\usepackage{url}
\usepackage{mathtools}
\usepackage{geometry}
\usepackage{hyperref}
\hypersetup{
    colorlinks=true,
    linkcolor=blue,
    filecolor=magenta,      
    urlcolor=blue,
    citecolor=blue,
    }
\usepackage{xcolor}
\usepackage{cite}
\usepackage{multirow}
\usepackage{multicol}
\usepackage{booktabs}
\usepackage{subfig}
\usepackage[T1]{fontenc}
\usepackage{lipsum}
\usepackage[utf8]{inputenc}
\usepackage{array}
\usepackage{lettrine}
\usepackage{authblk}
\usepackage{graphicx}
\usepackage{stfloats}
\usepackage{amsmath}
\usepackage{epsfig}
\usepackage{amssymb}
\usepackage{comment}
\usepackage[utf8]{inputenc}
\usepackage[mathletters]{ucs}
\usepackage{tablefootnote}
\usepackage{algcompatible}
\usepackage{algorithm}
\usepackage{algpseudocode}
\usepackage{setspace}
\usepackage{bbm}
\usepackage[english]{babel}
\addto\captionsenglish{\renewcommand{\figurename}{Fig.}}
\usepackage{amsthm}
\usepackage{balance}

\usepackage[skip=2pt,font=footnotesize]{caption}

\begin{document}
   \title{ Beyond the Limits of Rigid Arrays: Flexible Intelligent Metasurfaces for Next-Generation Wireless Networks} 
    	\newgeometry {top=25.4mm,left=19.1mm, right= 19.1mm,bottom =19.1mm}%
\author{Ahmed Magbool,~\IEEEmembership{Graduate Student Member,~IEEE,} 
Vaibhav Kumar,~\IEEEmembership{Member,~IEEE,} \vspace{-0.25cm}
\\ Marco Di Renzo,~\IEEEmembership{Fellow,~IEEE,} and Mark F. Flanagan,~\IEEEmembership{Senior Member,~IEEE}\thanks{Ahmed Magbool and Mark F. Flanagan are with the School of Electrical and Electronic Engineering, University College Dublin, D04 V1W8, Ireland (e-mail: ahmed.magbool@ucdconnect.ie, mark.flanagan@ieee.org). \par
Vaibhav Kumar is with Engineering Division, New York University Abu Dhabi (NYUAD), Abu Dhabi 129188, UAE (e-mail: vaibhav.kumar@ieee.org).\par 
Marco Di Renzo is with Universit\'e Paris-Saclay, CNRS, CentraleSup\'elec, Laboratoire des Signaux et Syst\`emes, 3 Rue Joliot-Curie, 91192 Gif-sur-Yvette, France (marco.di-renzo@universite-paris-saclay.fr), and with King's College London, Centre for Telecommunications Research -- Department of Engineering, WC2R 2LS London, United Kingdom (marco.di\_renzo@kcl.ac.uk). \par
The work of A. Magbool and M. F. Flanagan was supported by Research Ireland under Grant Number 13/RC/2077\_P2 and under Grant Number 24/FFP-P/12895. \par
The work of M. Di Renzo was supported in part by the European Union through the Horizon Europe project COVER under grant agreement number 101086228, the Horizon Europe project UNITE under grant agreement number 101129618, the Horizon Europe project INSTINCT under grant agreement number 101139161, and the Horizon Europe project TWIN6G under grant agreement number 101182794, as well as by the Agence Nationale de la Recherche (ANR) through the France 2030 project ANR-PEPR Networks of the Future under grant agreement NF-YACARI 22-PEFT-0005, and by the CHIST-ERA project PASSIONATE under grant agreements CHIST-ERA-22-WAI-04 and ANR-23-CHR4-0003-01. Also, the work of M. Di Renzo was supported in part by the Engineering and Physical Sciences Research Council (EPSRC), part of UK Research and Innovation, and the UK Department of Science, Innovation and Technology through the CHEDDAR Telecom Hub under grant EP/X040518/1 and grant EP/Y037421/1, and through the HASC Telecom Hub under grant EP/X040569/1.  }}

\maketitle

\begin{abstract}
Following recent advances in flexible electronics and programmable metasurfaces, flexible intelligent metasurfaces (FIMs) have emerged as {a promising enabling technology for next-generation wireless networks}. A FIM is {a morphable electromagnetic surface capable of dynamically adjusting its physical geometry to influence the radiation and propagation of electromagnetic waves}. Unlike conventional rigid arrays, FIMs introduce {an additional spatial degree of design freedom enabled by mechanical flexibility}, which can enhance beamforming, spatial focusing, and adaptation to dynamic wireless environments. {This added capability enables wireless systems to shape the propagation environment not only through electromagnetic tuning but also through controllable geometric reconfiguration}. This article explores the potential of FIMs for next-generation wireless networks. We first introduce {the main hardware architectures of FIMs and explain how they can be integrated into wireless communication systems}. We then present {representative application scenarios}, highlighting the advantages of FIMs for future wireless networks and comparing them with {other emerging flexible wireless technologies}. To illustrate their potential impact, we present case studies comparing FIM-enabled architectures with conventional rigid-array systems, demonstrating {the performance gains enabled by surface flexibility for both communication and sensing applications}. Finally, we discuss {key opportunities, practical challenges, and open research directions that must be addressed to fully realize the potential of FIM technology in future wireless communication systems}.
\end{abstract}

\IEEEpeerreviewmaketitle
\section{Introduction} 
Massive multiple-input multiple-output (MIMO) has emerged as a key technology for fifth-generation (5G) wireless networks, leveraging large antenna arrays to achieve high spectral efficiency, improved diversity gain, and enhanced spatial multiplexing. {However, future wireless systems are expected to support extremely high data rates and ultra-dense deployments, which require even finer spatial resolution.} Therefore, further scaling of antenna arrays becomes essential to provide finer spatial resolution. {This trend significantly increases hardware complexity and power consumption, motivating the exploration of alternative wireless architectures and enabling technologies.}

In response, metasurface-based solutions have gained significant attention. Programmable intelligent metasurfaces can manipulate electromagnetic (EM) waves through engineered, low-cost sub-wavelength elements, enabling controllable reflection, refraction, and focusing~\cite{2025_Magbool}. {By appropriately tuning the responses of these elements, metasurfaces can operate as transmitters, receivers, or reflectors to dynamically control how wireless signals propagate through the environment.} Consequently, intelligent metasurfaces can enhance coverage, improve signal quality, and reduce transmit power requirements.

Nevertheless, {The structural rigidity of most existing metasurface architectures limits the spatial adaptability of the surface and restricts the level of control that can be achieved over EM wave propagation.} Flexible intelligent metasurfaces (FIMs) introduce a transformative concept by enabling physical morphability of the EM surface~\cite{2025_An}. {In addition to electronic programmability, FIMs allow the geometry of the surface itself to be dynamically reconfigured, providing an additional spatial design dimension for wireless systems.} {This flexibility enables more precise control of the generated EM wave field, supporting improved beamforming, spatial focusing, and adaptive wavefront shaping.}

This {article} discusses the potential of FIMs {for} future wireless networks. In Section~\ref{sec:wor_prn}, we first introduce the hardware architectures of FIMs along with their utilization in wireless networks. Section~\ref{sec:emrg_trnds} highlights key application scenarios and compares FIMs with other relevant flexible architectures. {Section~\ref{sec:cs_std} presents representative case studies that illustrate the performance advantages of FIM-enabled communication systems over conventional rigid arrays.} Finally, Section~\ref{sec:opp_chl} discusses opportunities, challenges, and open research directions, and Section~\ref{sec:conc} concludes the {article}.

\section{Hardware Architectures of FIMs and Their Utilization in Wireless Networks} \label{sec:wor_prn}

In this section, we introduce the hardware architectures of FIMs from an electronic perspective. We also discuss how FIMs can be integrated into wireless networks.

\begin{table*}
\footnotesize
\centering
\caption{Hardware classification of FIMs.}
\begin{tabular} {|m{1.8cm}|m{2.5cm} | m{1.8cm} |  m{3cm} | m{3cm} | m{3cm} |} 
 \hline Hardware architecture & Morphing mechanism & Materials & Shape reconfiguration & Key advantages & Main limitations \\  [0.5ex] 
 \hline 
 \multicolumn{6}{c}{Passive surfaces}\\
 \hline
  Stretchable metasurfaces & External in-plane stretching or compression modifies inter-element spacing and effective aperture geometry & PDMS, elastomeric polymers & Smooth geometric deformation dictated by applied mechanical strain; no embedded actuation or feedback & Extremely low hardware complexity; zero power consumption for shape control; high mechanical robustness & Geometry cannot be reconfigured on demand; deformation depends on external forces and lacks spatial selectivity \\
  \hline
  Conformal dielectric metasurface & Surface bends to follow curved host objects or structures & Polyimide, PDMS & Global curvature fixed by mounting surface; deformation remains static during operation & Simple integration on curved platforms; preserves phase control under moderate bending & No dynamic or programmable shape adaptation; limited adaptability to changing environments \\
    \hline
  Substrate-compliant meta-atom arrays & Elastic substrates allow small passive displacement of meta-atoms under mechanical loading & Soft polymers & Bounded perturbation of element positions within elastic limits & Stable EM response; minimal control overhead & Limited deformation range; geometry cannot be actively optimized \\
  \hline
  \multicolumn{6}{c}{Active surfaces}\\
 \hline
 Liquid-metal microfluidic FIMs & Electrically or pressure-driven redistribution of liquid metal in embedded microchannels & PDMS with microfluidic networks & Localized, continuous, and reversible deformation controlled via electrical or pneumatic inputs & Fine-grained and smooth 3D geometry control; enables joint EM–mechanical optimization & Fabrication complexity; sealing, leakage, and long-term reliability issues \\
 \hline
 Lorentz-force actuated FIMs & Electrical currents interacting with magnetic fields generate controllable out-of-plane forces & Elastomers with conductive meshes & Fast and spatially programmable deformation enabling dynamic 3D surface shaping & High reconfiguration speed; enables software-defined aperture shaping & Elevated power consumption; complex control circuitry; potential thermal issues \\
 \hline
 Soft electro-mechanical FIMs & Embedded mechanical or electrothermal actuators induce controlled displacement & Elastomers, flexible printed circuit board (PCBs) & Programmable, repeatable deformation patterns with discrete or quasi-continuous resolution & High precision and repeatability; decoupled control of EM tuning and geometry & Actuator fatigue; scalability challenges for large apertures \\
 \hline
\end{tabular}
\label{table:1}
\end{table*}

\begin{figure*} 
         \centering
\includegraphics[width=2\columnwidth]{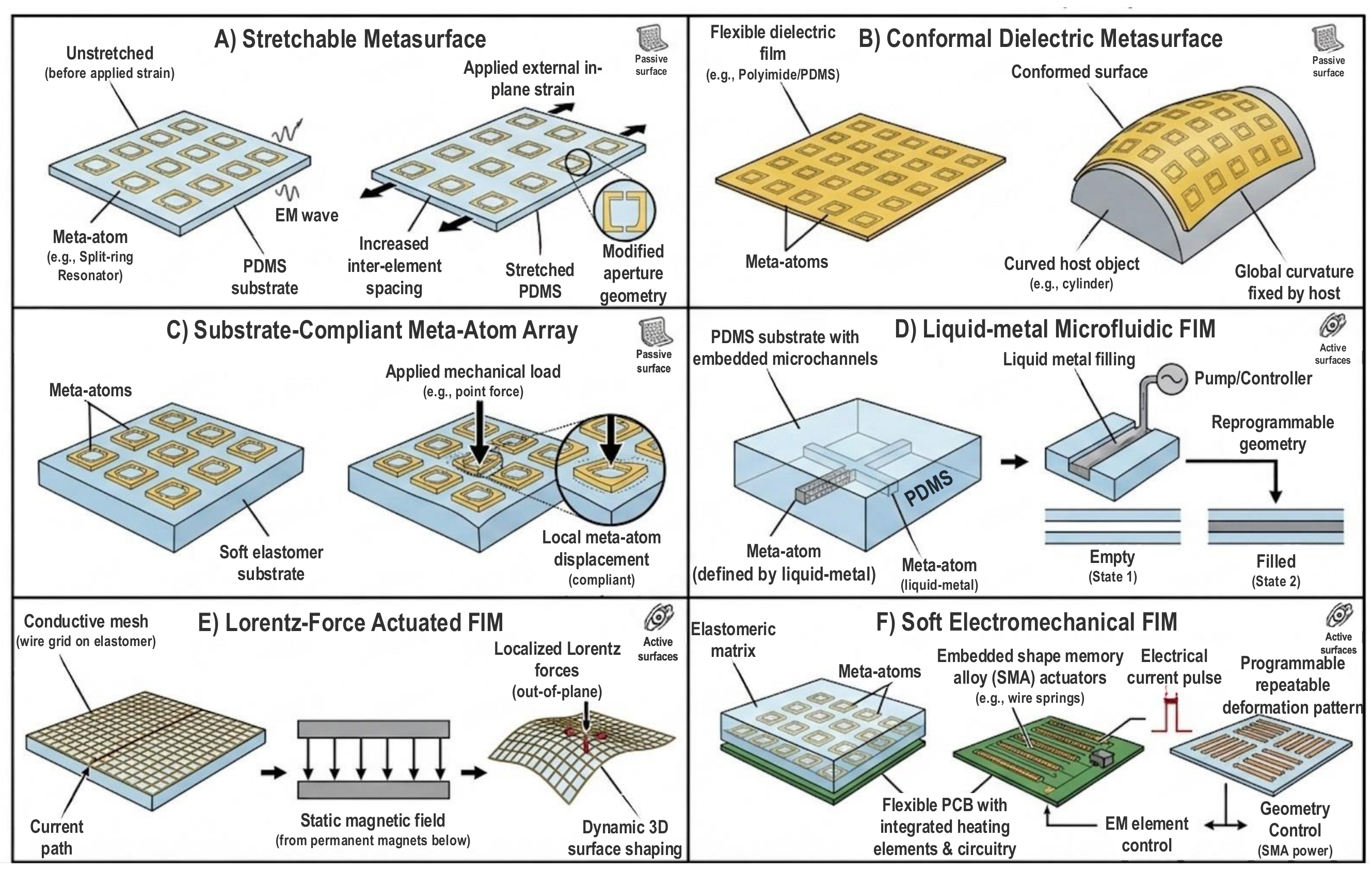}
        \caption{Hardware architectures of FIMs.}
        \label{fig:main_fig}
\end{figure*}

\subsection{Hardware Architectures of FIMs}

FIMs extend conventional reconfigurable surfaces by introducing controlled geometric adaptability in addition to EM programmability. Unlike rigid metasurfaces with fixed element locations, FIM platforms are implemented on mechanically compliant substrates that allow the aperture to deform while maintaining {the desired} EM behavior. Recent studies have demonstrated that flexible metasurface implementations can preserve wave manipulation capability even under significant bending, confirming their suitability for conformal wireless environments~\cite{li2025flexible}.

As shown in Table~\ref{table:1} and Fig.~\ref{fig:main_fig}, FIM architectures can be broadly categorized into \emph{passive} surface morphing and \emph{active} surface morphing. Passive morphing designs rely primarily on intrinsic material flexibility, where the surface shape changes due to external mechanical loading {such as bending or stretching}. Typical implementations employ elastomeric substrates including polyimide and polydimethylsiloxane (PDMS), which provide {low mechanical stiffness and favorable EM properties}. For instance, dielectric metasurface inclusions embedded in PDMS have demonstrated efficient conformal wave manipulation while maintaining phase control capability~\cite{cheng2016all}. These approaches offer structural simplicity and low power consumption but provide limited {capability for real-time} shape control.

Active morphing FIMs incorporate embedded actuation mechanisms that enable programmable and localized deformation of the surface. Representative implementations include liquid-metal microfluidic networks and Lorentz-force driven conductive meshes, where spatially distributed electrical currents generate controllable out-of-plane displacement. {These platforms can achieve fast, reversible, and continuous transformations between complex three-dimensional surface profiles, enabling \emph{software-defined aperture shaping}~\cite{ni2022soft}.} While active morphing provides significantly higher adaptability, it introduces additional design challenges in terms of power consumption, control circuitry, and mechanical reliability.

\subsection{FIM Utilization in Wireless Networks}

\begin{figure*} 
         \centering
\includegraphics[width=1\linewidth]{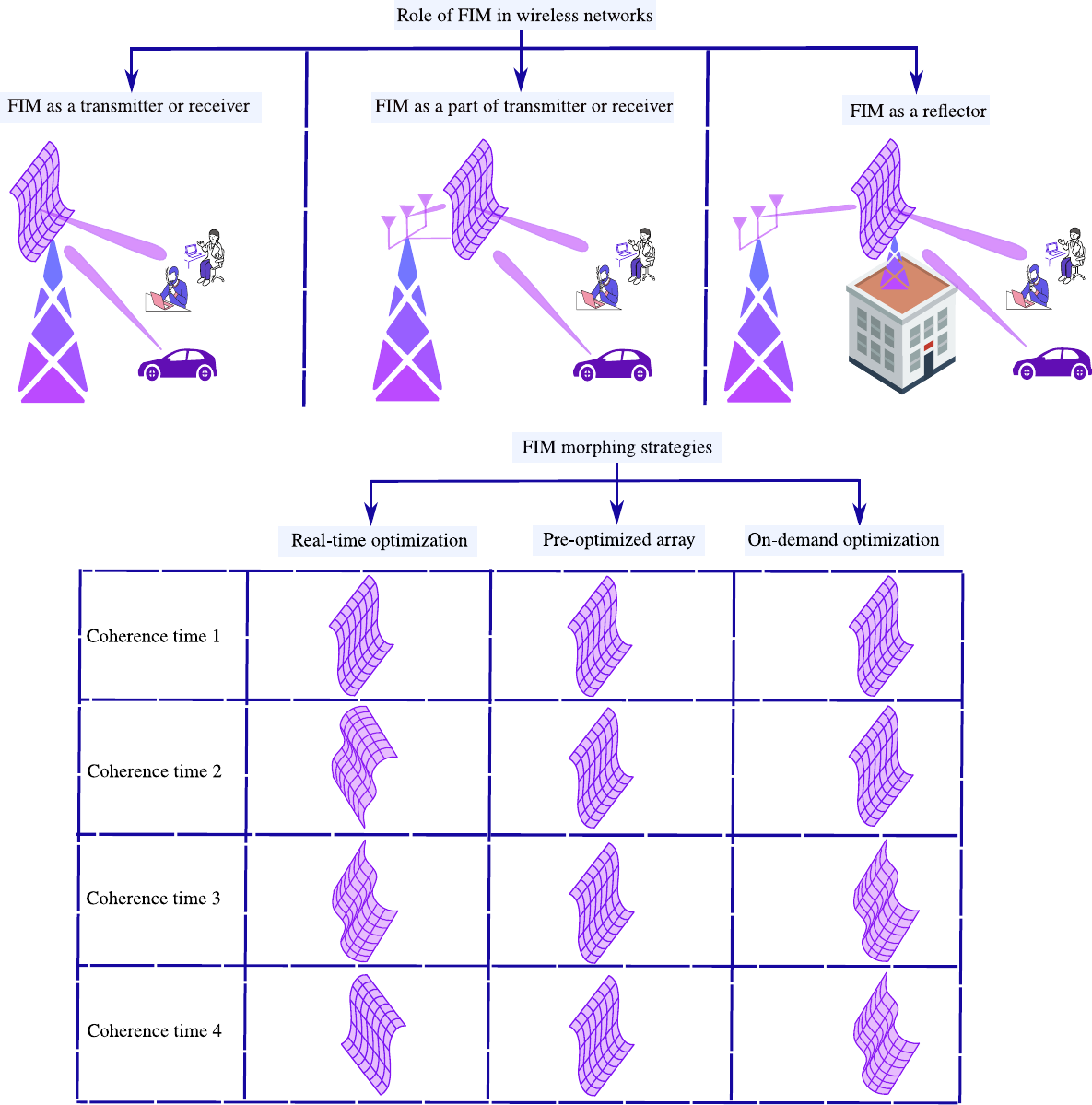}
        \caption{{Role of FIM} in wireless networks and morphing strategies.}
        \label{fig:FIM_mor_str}
\end{figure*}

As shown in Fig.~\ref{fig:FIM_mor_str}, FIMs can directly function as transmitters and/or receivers in wireless networks. When used as transmitters, FIMs can generate tailored radiation patterns by dynamically adjusting their geometry. Similarly, when operating as receivers, FIMs can adapt their surface shape to focus incoming signals, thereby improving signal quality. Compared to rigid arrays, FIM-based systems can achieve the same channel capacity with a substantially lower transmit power budget. For example,~\cite{2025_An} {shows} that a capacity of \unit[3]{bit/s/Hz} {can be achieved} with \unit[6.5]{dBm} transmit power using FIMs, whereas a rigid array requires \unit[12.5]{dBm} for the same performance. Furthermore,~\cite{2025_An1} demonstrated that the inherent selection gain of FIMs can reduce the required transmit power for multi-user communication by nearly half compared to conventional rigid arrays under identical data rate constraints. It should be noted, however, these studies are simulation-based and focus solely on the transmit power budget at the base station, without accounting for the additional power required to physically morph the FIM itself.

FIMs can also be integrated into the transmitter architecture to enhance overall system performance by employing a FIM layer to shape the outgoing wavefront. This configuration differs from using the FIM as a transmitter directly, since conventional transmitting arrays are still required. The same approach can be applied at the receiver. This hybrid architecture enables joint optimization between digital baseband and spatial signal processing, offering an effective trade-off between performance and hardware efficiency.

FIMs can also be utilized as reflectors placed between transmitters and receivers. Unlike conventional rigid RISs, FIM reflectors can physically deform to adapt to environmental changes, user mobility, or coverage requirements. This morphing capability enables more flexible and efficient wireless environments, providing enhanced coverage, reduced signal blockage, and improved spatial diversity in dynamic scenarios. For example, it was demonstrated in~\cite{2025_Hu} that a FIM can achieve a \unit[3]{dB} improvement in channel gain compared to {a rigid reconfigurable intelligent surface (RIS)-assisted system}.

\subsection{FIM Morphing Strategies}

{FIM morphing can follow different strategies depending on how frequently the surface geometry is updated in response to channel variations, as illustrated in Fig.~\ref{fig:FIM_mor_str}. In practice, the choice of morphing strategy determines the trade-off between system performance, control overhead, and implementation complexity.}

The approach which has received most attention in the literature is that of real-time morphing, in which the FIM continuously adapts its configuration according to the instantaneous channel conditions. In this strategy, the surface response is optimized within the channel coherence time, enabling the system to react to rapid fluctuations in user locations, mobility, and small-scale fading. {While this strategy can provide the best instantaneous performance, it requires fast control circuitry, low-latency network feedback, and increased computational resources to determine the optimal surface configuration.}

The second morphing strategy is pre-optimized static morphing, in which the surface is configured based on long-term or statistical channel information. In this case, the surface shape is synthesized once according to these pre-determined settings. Thus, these surfaces are not considered flexible in the strict sense. This approach is suitable for typical user distributions or average propagation conditions. Pre-optimized morphing significantly reduces control overhead, simplifies hardware requirements, and avoids the need for network feedback. {However, since the surface configuration remains fixed during operation, the system cannot adapt to instantaneous channel fluctuations or user mobility.}

{A third strategy is a hybrid approach that combines both real-time and pre-optimized morphing. In this case, the FIM is initially configured using long-term channel statistics, while limited real-time adjustments are applied to compensate for instantaneous channel variations. Only when the propagation environment changes significantly, such as due to user relocation or changes in the surrounding environment, is the statistical configuration recomputed. This hybrid strategy offers a practical compromise between performance and complexity, enabling adaptive operation without requiring continuous real-time reconfiguration.}

\section{Emerging Applications of FIMs in Wireless Networks} \label{sec:emrg_trnds}

\begin{figure*} 
         \centering
\includegraphics[width=0.8\linewidth]{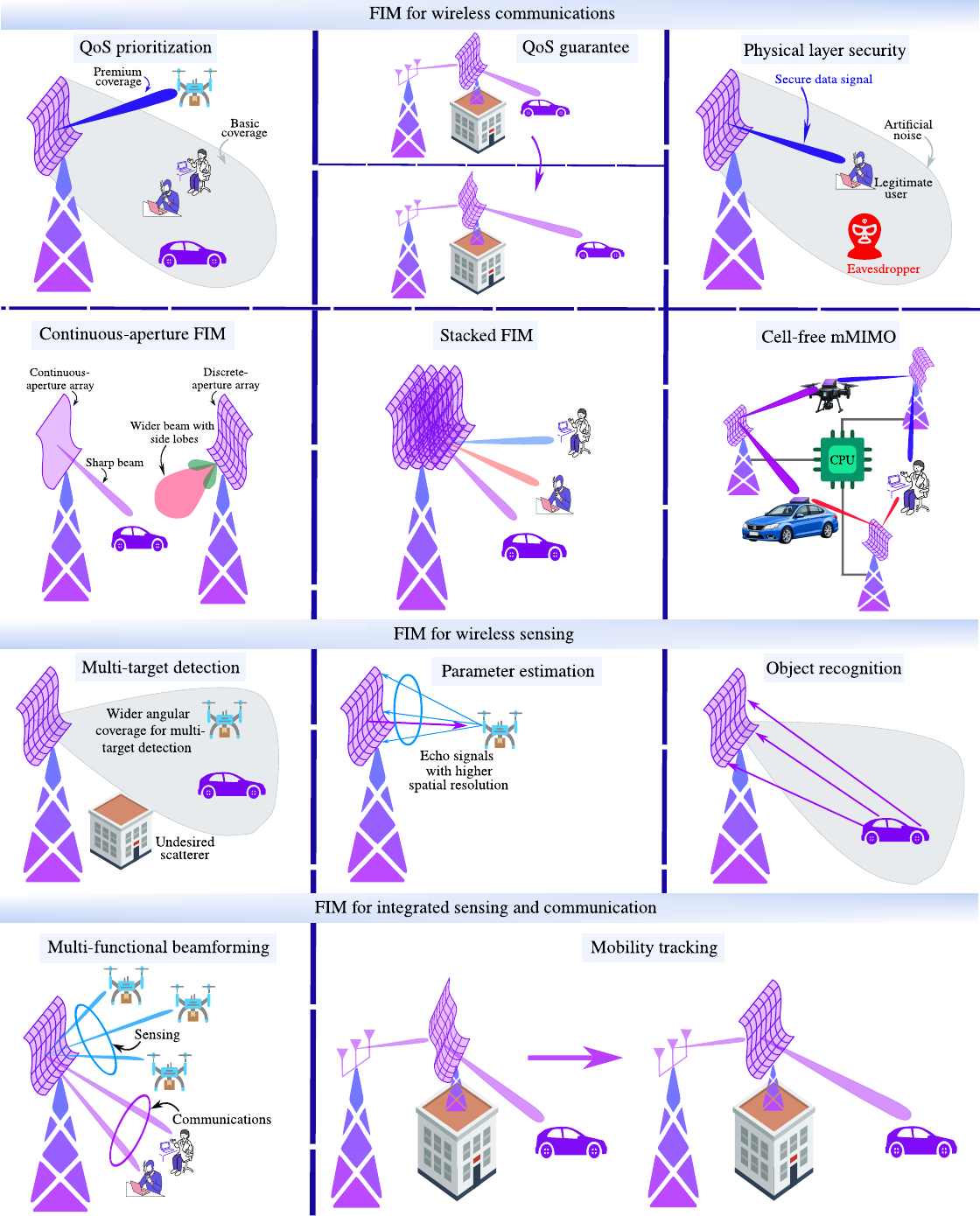}
        \caption{Potential applications of FIMs in wireless networks.}
        \label{fig:FIM_app}
\end{figure*}
 
Fig.~\ref{fig:FIM_app} illustrates several emerging applications where FIMs can be leveraged to enhance the performance of wireless networks. These applications span communication, sensing, and integrated sensing and communication (ISAC).

\subsection{FIM for Wireless Communications}

{From a system perspective, the benefits of FIMs for wireless communications can be realized across a wide range of categories.}

FIMs can be used as an architectural enabler for next-generation communications across a wide range of scenarios. One of these is the use of FIM to enable user prioritization by dynamically morphing the surface response to create user-dependent propagation conditions. This allows the network to prioritize users according to their service requirements while still providing a basic service level to others. For instance, consider a system serving multiple users with heterogeneous quality-of-service (QoS) requirements, where some users require high throughput or enhanced reliability (e.g., premium or latency-sensitive users), while others only need baseline connectivity. The surface can be morphed to synthesize sharp focal spots or directive beams that deliver stronger constructive interference at the locations of the high-priority users. Simultaneously, a broader or diffused radiation pattern can be maintained to ensure coverage continuity for the remaining users. While rigid-array-based systems can also provide this multi-user prioritization, FIMs offer greater flexibility, as morphing enables reinforcement of strong paths to priority users while suppressing interfering paths.

Moreover, FIMs can address the challenge of maintaining consistent QoS in cellular systems despite rapid channel variations caused by user mobility and small-scale fading. By continuously adjusting the surface morphing, the reflected or transmitted wavefront can be reshaped to reinforce the desired signal paths and compensate for channel degradation, thereby stabilizing the effective channel experienced by the user. This capability is illustrated in~\cite{2025_An1}, where the authors demonstrate that, in a single-user scenario, employing a FIM as a transmitter with a morphing range of $0.8$ of the wavelength can achieve a 100\% increase in received signal power compared with {a rigid-array-based system}. This indicates that FIM-based systems can provide more adaptive coverage, particularly for users under challenging propagation conditions such as cell-edge locations or severe fading environments.

Furthermore, FIMs can serve as enablers for physical layer security (PLS). Through programmable surface morphing, the propagation of confidential signals can be spatially shaped to concentrate energy toward legitimate users while diffusing, scattering, or decorrelating the signal in unintended directions. This spatial selectivity reduces the likelihood that an eavesdropper observes a strong or coherent version of the transmitted waveform. For example, the surface can be configured to focus the useful signal toward the legitimate user while simultaneously synthesizing artificial noise in other spatial directions. This can be achieved by programming subsets of the surface elements to generate spatially distributed interference patterns outside the intended coverage region. Consequently, while the authorized user receives a coherent and high-quality signal, potential eavesdroppers observe a mixture of weak signal components and artificial noise, significantly degrading their decoding capability.

In addition, by densely packing FIMs’ meta-atoms, a flexible continuous aperture array surface can be realized. These surfaces can generate holographic beam patterns for precise beamforming, thereby improving system performance. A simulation-based study in~\cite{2025_Ranasinghe} demonstrated that at a morphing range equal to the wavelength, the average user rate can increase by 49\% compared to a conventional discrete FIM-based system and by 30\% compared to a rigid continuous-aperture-array-based system.

FIMs can also be integrated with other emerging technologies to overcome inherent limitations and achieve enhanced performance gains. One such approach involves employing FIMs as layers in stacked intelligent metasurfaces (SIMs), as investigated in~\cite{2026_Magbool}. The integration of SIM and FIM architectures enables enhanced wave-domain processing gains. The results reported in~\cite{2026_Magbool} demonstrate that this integration mitigates the performance saturation observed with an increasing number of SIM layers. Consequently, it facilitates more efficient wave-domain signal processing capabilities compared to rigid SIM-based systems.

Another compelling direction is the use of FIMs as passive reflectors in cell-free massive MIMO systems. In this case, FIMs can help reduce the number of access points required while improving coverage in areas that rigid RISs cannot not reach, enhancing both user rates and overall system efficiency.

\subsection{FIM for Wireless Sensing}
{FIMs can support wireless sensing across the three main sensing tasks: target detection, parameter estimation, and object recognition.} 

In target detection, FIMs can enhance multi-target detection by adjusting their morphing to form multiple steerable beams or focal regions that illuminate distinct spatial sectors simultaneously. This capability allows FIM-based systems to probe a wide area efficiently and detect the presence of several objects within a single observation period. {The physical morphing of the surface can be viewed as a low-complexity analog beam scanner that extends the field of view and improves the probability of detecting weak or partially obscured targets without relying on extensive digital beamforming hardware.} FIMs can also mitigate the negative effects of undesirable random non-line-of-sight (NLoS) components, which would otherwise degrade sensing accuracy. {The authors of~\cite{2025_Teng} showed that a FIM-based sensing system can improve the received sensing power by 44.5\% compared to a rigid-array-based sensing system.}

FIMs can further support parameter estimation by dynamically refining the beam patterns to improve resolution in range, angle, and velocity domains. Through continuous shape adaptation, the surface can narrow its beamwidth to isolate individual echoes from closely spaced targets or broaden it to cover {a} large angular sector. Moreover, the morphing geometry can be optimized to diversify the effective phase distribution across the surface, enhancing the precision of delay and Doppler estimation. In addition, by adjusting the effective propagation path or virtual aperture through {surface} morphing, FIMs can improve the estimation accuracy for target parameters. 

FIMs can also facilitate target recognition by introducing controlled diversity in the scattering environment beyond what is achievable with conventional large{-scale} MIMO arrays. While MIMO systems can generate angular diversity through beam steering, FIMs provide an additional degree of control by physically reshaping the propagation environment itself. By sequentially altering their morphing states, FIMs induce distinct scattering configurations, enabling the capture of backscattered signals under different boundary conditions. This multi-configuration sensing enriches the acquired data with complementary spatial and spectral features, improving target distinguishability.

\subsection{FIM for ISAC}

Since FIMs can support both communication and sensing tasks, their utilization for ISAC can provide benefits to both functionalities while also improving the associated trade-offs. A key design challenge in ISAC systems is achieving an effective balance between communication and sensing performance. Since element morphing introduces an additional degree of control that can be exploited, FIM-enabled systems can achieve a superior Pareto-optimal front compared to conventional rigid-array-based systems. Moreover, FIMs can form sharper and more adaptive wavefronts, allowing the surface to allocate its aperture more efficiently across multiple users and sensing targets. This results in improved sensing accuracy without significantly sacrificing communication throughput, or vice versa.

Moreover, FIMs dynamically adjust surface curvature to synthesize directional beams that adapt to rapid changes in user position or target motion. Such spatiotemporal control enables continuous alignment of the communication link while maintaining target illumination, effectively compensating for Doppler shifts and time-varying channel states in vehicular or unmanned aerial vehicle (UAV) networks~\cite{2025_Kuranage}. {In such scenarios, a simple morphing strategy can be employed based on radar-assisted predictive beamforming, following principles similar to those in~\cite{2020_Liu}.} This type of approach leverages real-time measurements together with state-evolution models to track channel variations, enabling fast channel acquisition. Based on these predictions, a low-complexity morphing configuration can then be selected to reconfigure the surface, adapt the propagation pattern, and help maintain reliable coverage for moving users.

\subsection{Other Flexible Antenna Technologies in Wireless Networks}
Several other flexible antenna technologies have recently been proposed for wireless systems\footnote{For more details on flexible antenna technologies in wireless networks, please refer to~\cite{2026_Wang} and the references therein.}. They can be broadly {grouped} into two {categories}.

The first category involves architectures in which the antenna positions within an array can be adjusted, commonly referred to as movable antennas. The movement of antennas can be realized through {various} mechanisms, including mechanical actuation, pinching, and fluid-based approaches. Movable antennas are particularly well suited for applications in which element-level repositioning is the main {performance-enabling mechanism}. {Typical examples include blockage mitigation in millimeter-wave (mmWave) links, where antennas can be physically relocated to restore line-of-sight paths. Another example is user tracking in dynamic environments, where antenna positions are adjusted to follow mobile users and preserve favorable channel conditions.} In contrast, FIMs are better suited to applications in which global surface geometry and aperture shaping are critical. {For example, in dense multipath environments, FIMs can continuously morph their surface to synthesize highly directive or focused beams, control wavefront curvature, and tailor the spatial field distribution.} This capability enables advanced functions such as simultaneous multi-user beamforming, spatial focusing for sensing, and enhanced interference management, which are difficult to achieve through antenna repositioning alone.

The second category comprises architectures in which the entire antenna surface, rather than individual elements, can be repositioned or reoriented. A prominent example is rotational arrays, in which the antenna surface is mechanically rotated to steer beams or reorient the array toward desired users or reflectors, without requiring active phase control at each element. Such arrays are well suited to scenarios requiring coarse angular adaptation. For instance, rotational arrays can be adjusted to illuminate specific angular sectors of interest. In contrast, FIMs are particularly effective in scenarios where continuous aperture shaping offers advantages over physically movable surfaces. {For example, in an urban street canyon, FIMs can continuously reshape the beam pattern at a sub-wavelength scale to focus energy toward users, without being limited to a finite set of angular sectors.}

\section{Case Studies} \label{sec:cs_std}
In this section, two representative case studies are presented to illustrate the performance benefits of FIM{-enabled systems} for communications and sensing compared with conventional rigid-array {systems}. Two systems are considered: (i) a downlink multi-user communication system with the objective of maximizing the system's sum rate, and (ii) a multi-target sensing system with the objective of maximizing the received power toward specified angles. Both systems operate at a carrier frequency of 28~GHz, corresponding to a wavelength of approximately $\lambda \approx \unit[10.7]{mm}$. In both cases, the total power budget is set to $\unit[25]{dBm}$.

\subsection{FIM-Enabled Communications}
In our first case study, a base station equipped with a FIM having $M$ elements ($M \in \{9, 16\}$) serves 8 users in the downlink, where the morphing range varies between {$0$} and $\lambda$. {A multipath channel model is considered considered, with I paths ($I \in \{4, 8\}$).} The receiver noise power is $\unit[-104]{dBm}$.

\begin{figure}
  \centering
  \begin{tabular}{|c|}
  \hline
    \includegraphics[width=0.9\columnwidth]{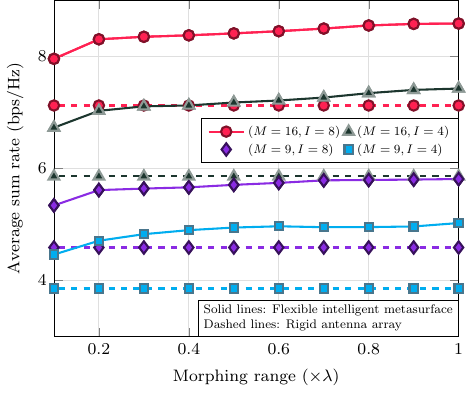} \\
    \scriptsize (a) \\
    \includegraphics[width=0.9\columnwidth]{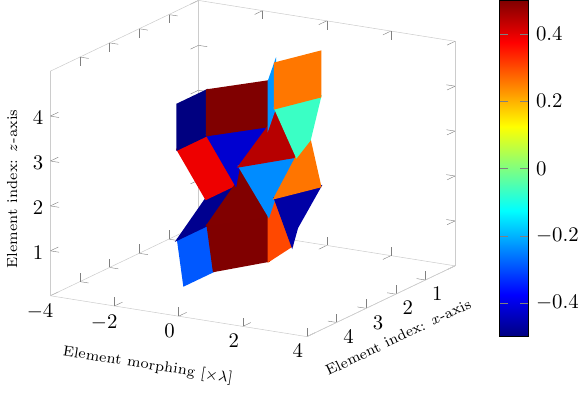} \\
     \scriptsize (b) \\
        \hline
  \end{tabular}
    \medskip
  \caption{(a) Average sum rate versus the normalized morphing range, and (b) Example optimized surface morphology for $M=16$, $I=8$ and a morphing range of $0.5 \lambda$.}\label{fig:comm}
\end{figure}

Fig.~\ref{fig:comm} shows the average sum rate as the allowable morphing range increases. Several important trends can be observed. First, the FIM-enabled system consistently outperforms the rigid-array baseline across all configurations. The performance gain grows rapidly in the small-morphing regime and gradually saturates beyond approximately $0.4\lambda$. This behavior indicates that most of the spatial diversity benefits can be captured with moderate structural deformation.

In addition, it can be observed that FIM{s are} more effective in sparser propagation environments, as {they can morph the surface to focus energy on strong paths while avoiding weak ones, which can be better controlled under such conditions}. For instance, the percentage of improvement as the morphing range increases from zero to $\lambda$ is 23\% and 28\% for eight and four propagation paths, respectively, using nine transmit antennas.

To provide physical insight, Fig.~\ref{fig:comm} also visualizes a representative optimized surface configuration. The resulting topology is clearly non-uniform, reflecting the strong coupling between the channel realization and the preferred aperture shape. Instead of maintaining a fixed planar geometry, the FIM selectively adjusts element positions to reshape the effective wavefront.

\subsection{FIM-Enabled Sensing}
In the FIM-enabled sensing system, we aim to form beams to detect three spatially distributed targets. The FIM has $M=9$ radiating elements and the morphing range is set to $\lambda$. A line-of-sight (LoS) channel model is considered, with the objective of maximizing the received signal power at three distinct angles.

\begin{figure}
  \centering
  \begin{tabular}{|c|}
  \hline
    \includegraphics[width=0.9\columnwidth]{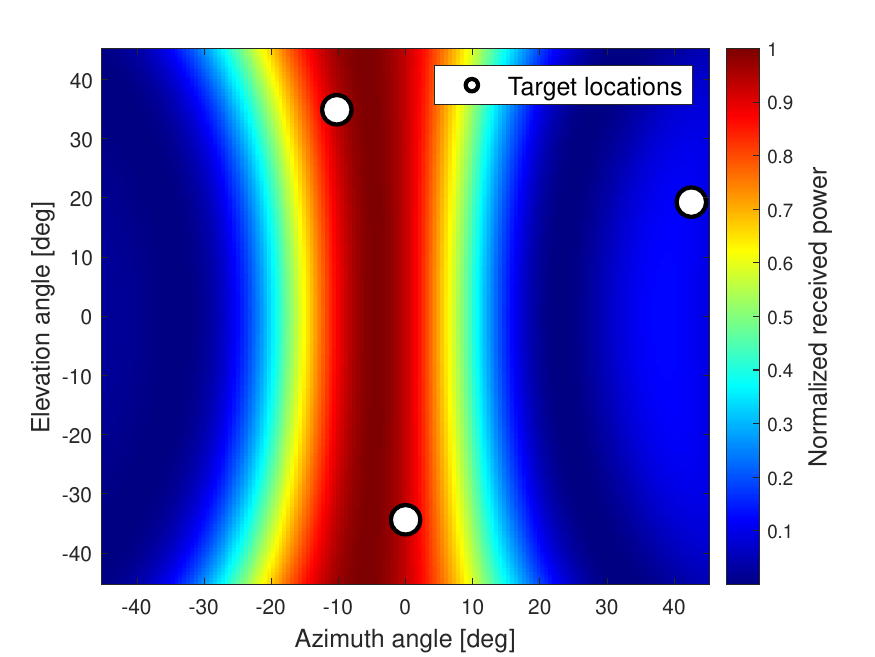} \\
    \scriptsize (a) \\
    \includegraphics[width=0.9\columnwidth]{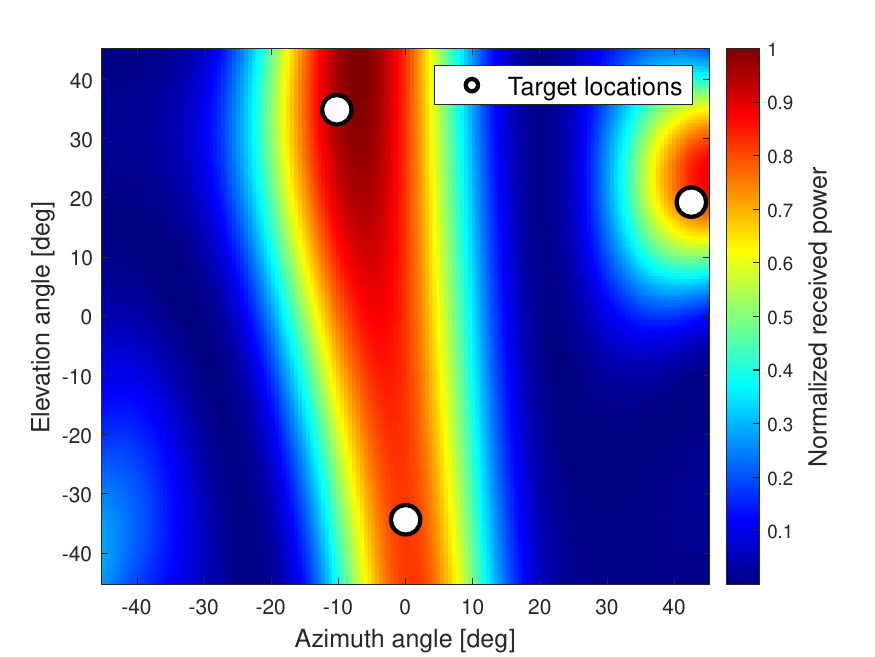} \\
     \scriptsize (b) \\
      \includegraphics[width=0.9\columnwidth]{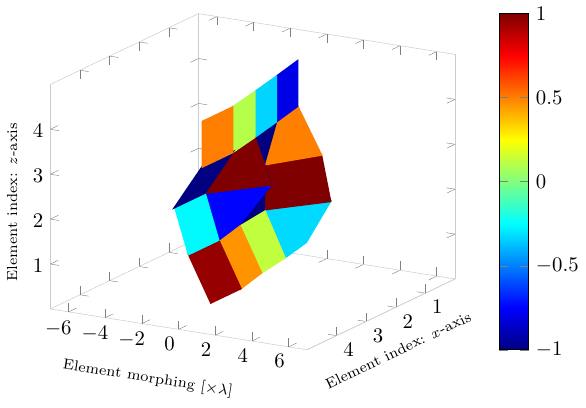} \\
       \scriptsize (c) \\
        \hline
  \end{tabular}
    \medskip
  \caption{Normalized received power for a system aiming to detect three targets using (a) a rigid, and (b) a FIM. In (c) the FIM response for target detection is shown.}\label{fig:sens}
\end{figure}

Fig.~\ref{fig:sens} depicts heatmaps of the normalized received power at each location, defined as the received power at the location normalized by the maximum received power across all locations. It can be observed that the rigid array has limited beamforming capability, as it can illuminate two targets but fails to reach the third. In contrast, the FIM-enabled system demonstrates a greater ability to form beams toward the three locations, resulting in enhanced target detection performance. This improvement is enabled by the non-trivial surface morphing, which introduces additional spatial degrees of freedom and allows more flexible wavefront shaping, thereby enabling simultaneous power focusing toward multiple angular directions.

\section{Opportunities, Challenges and Future Research Directions} \label{sec:opp_chl}
Research on employing FIMs in wireless networks is still at an early stage, and further investigation is needed to fully assess their practical potential. This section discusses the key opportunities, challenges, and future research directions for leveraging FIMs in next-generation wireless systems.
\subsection{Channel Modeling and Estimation}
Accurate channel modeling is important for system optimization, as it influences estimation performance, resource allocation, and overall network efficiency. Most studies on FIM-assisted systems adopt simplified multipath channel models that are suitable for mmWave and THz bands due to their sparse scattering environments. However, these models may not capture the dense scattering and complex propagation effects present in lower frequency bands, where rich multipath components dominate.

Once an appropriate channel model is available, reliable channel estimation algorithms are needed to exploit the potential of FIM-assisted networks. The flexible architecture of FIMs can capture richer spatial and angular information about the propagation environment. However, the additional control variables introduced by the morphable structure increase the dimensionality and complexity of the estimation problem, as channel parameters may be coupled with surface configuration variables, leading to a nonlinear estimation task. At higher frequencies, such as mmWave and THz, propagation environments typically contain fewer dominant paths, which can reduce pilot overhead. These structural properties can be leveraged in algorithm design. Existing works have explored channel estimation schemes that jointly consider the EM behavior of FIMs and communication objectives using techniques such as sparse recovery~\cite{2025_Jiang} and neural network–based approaches~\cite{2026_Xiao}.

\subsection{Energy Efficiency}
Next-generation wireless networks require energy-efficient system designs. The reduced number of active radio frequency (RF) chains in FIM-assisted systems can improve system-level energy efficiency. In addition, FIMs can enhance beamforming gain and spatial directivity without increasing transmit power, enabling target communication QoS to be achieved with lower transmit power than in conventional rigid-array systems. This capability enables energy-aware transmission strategies where surface configuration and power allocation are jointly optimized.

However, accurately quantifying the energy efficiency of FIM-assisted systems remains an open problem. In particular, experimentally validated power consumption models for surface morphing are still lacking. The mechanisms responsible for changing the surface shape may introduce non-negligible power overhead, depending on morphing range, reconfiguration speed, and control circuitry complexity. Furthermore, frequent reconfiguration to track time-varying channels could offset transmit power savings. Therefore, comprehensive power consumption models that include transmit power, RF operation, control signaling, and morphing energy costs are required to properly evaluate the net energy benefits of FIMs.

\subsection{Hardware Design and Morphing Speed}
Although FIMs promise significant performance gains through dynamic reconfiguration, their realization is constrained by hardware feasibility and achievable morphing speed. If the surface can be reconfigured faster than the channel coherence time, the system can adapt its geometry to instantaneous channel conditions, maximizing beamforming gain, spatial multiplexing, and interference suppression. This is particularly beneficial in low-mobility or quasi-static environments with relatively long coherence times. Even when real-time morphing is not possible, FIMs can still be optimized on a slower timescale based on large-scale channel statistics such as path loss, angular spreads, and dominant propagation directions.

However, whether the surface can morph fast enough to track small-scale fading remains uncertain. If morphing speed is slower than the channel coherence time, instantaneous channel-aware designs may become impractical. In addition, repeated rapid morphing may introduce mechanical fatigue, reliability issues, and increased power consumption. Therefore, the trade-off between morphing speed, structural stability, energy consumption, and system performance requires careful investigation.

Practical FIM-assisted systems must also account for hardware imperfections. Quantized morphing levels may limit surface adjustment resolution, reducing beamforming accuracy and spatial focusing capability. Moreover, hardware impairments such as non-ideal phase responses, metasurface losses, mechanical wear, and calibration errors can degrade performance. These constraints highlight the need to jointly consider hardware limitations, reconfiguration capabilities, and communication objectives when designing FIM-assisted wireless systems.

\subsection{Machine Learning}
Learning-based methods can help address challenges related to channel estimation, resource allocation, hardware impairments, and energy-aware adaptation in FIM-enabled networks, where model-based solutions may become analytically intractable.

For instance, in stacked FIM architectures shown in~\cite{2026_Magbool} (i.e., when multiple morphable layers are cascaded), the overall system can emulate the structure of a neural network. The propagation of EM waves through successive layers resembles the forward pass of a deep model, while the morphable configurations act as adjustable parameters. Unlike rigid SIM architectures, where the response is typically linear or limited to phase adjustments, stacked FIM layers can introduce richer nonlinear transformations. Effective nonlinear behavior may emerge due to geometry-dependent EM interactions, material responses, and inter-layer coupling, enabling the surface to approximate complex input{-}output mappings.

From this perspective, the stacked FIM itself has the potential to perform learning and signal processing directly in the physical domain. By adjusting the configurations of the morphable layers based on observed data, the surface can progressively improve its response to the propagation environment. This capability can support tasks such as adaptive beamforming, interference mitigation, and channel estimation. Such approaches can reduce the reliance on explicit analytical channel models and instead exploit data-driven adaptation based on observed channel conditions.

Machine learning can also play an important role in channel estimation for FIM-assisted networks. Due to the morphable geometry of the surface and the coupling between the channel parameters and the surface configuration, deriving accurate analytical channel models and estimation procedures can be challenging. Data-driven approaches can instead learn the relationship between pilot observations, surface configurations, and the underlying channel state. For example, neural networks can be trained to infer channel parameters or effective channel responses from a limited number of pilot measurements, potentially reducing pilot overhead. Such methods can also incorporate prior structural information about the propagation environment, such as spatial sparsity at {mmWave} and terahertz frequencies, enabling more efficient channel reconstruction even in high-dimensional FIM configurations.

Despite these opportunities, several challenges remain. The learning and adaptation process must account for physical constraints, quantized morphing levels, hardware imperfections, and energy limitations. Acquiring sufficiently rich and representative data can also be difficult, particularly in dynamic wireless environments. Moreover, enabling efficient online adaptation requires reliable feedback mechanisms and low-latency control of the morphable layers remains an important challenge for FIM-enabled wireless systems.


\section{Conclusion} \label{sec:conc}
{Flexible intelligent metasurfaces (FIMs) introduce a new degree of freedom for wireless system design by combining EM programmability with surface morphability. In this article, we discussed how this capability can help overcome key limitations of conventional rigid architectures and enable new opportunities for future wireless networks. We presented the hardware architectures of FIMs, described their roles in wireless communication and sensing systems, and highlighted emerging applications in communication, sensing, and ISAC. We also discussed representative case studies that illustrate the performance gains enabled by surface flexibility, and we outlined key opportunities, technical challenges, and future research directions. Overall, FIMs represent a promising architectural enabler for next-generation wireless systems, although their practical realization will depend on advances in channel modeling, hardware design, control strategies, and energy-efficient implementation.}



%





\ifCLASSOPTIONcaptionsoff
  \newpage
\fi





\bibliographystyle{IEEEtran}
\bibliography{IEEEabrv,Bibliography}
%


\end{document}